\documentstyle[11pt,appb,epsfig]{article}
\begin{document}
\title{Configuration Mixing Effects in Isoscalar Giant Dipole Resonance
\thanks{In honor of Professor Josef Speth on his 60th birthday.}
}
\author{M. W\'ojcik and S. Dro\.zd\.z
\address{ Institute of Nuclear Physics, PL-31-342 Krak\'ow, Poland\\
Institut f\"ur Kernphysik, Forschungszentrum J\"ulich,\\
D-52425 J\"ulich, Germany }
}
\maketitle
\begin{abstract}
Based on an explicit verification of the coupling matrix elements 
between the 1p1h and 2p2h states we propose a new 
method of selecting the most important 2p2h states responsible for 
fragmentation effects. In this way the dimensionality of the problem
is reduced, such that the computation becomes feasible 
and the spreading of the strength is realistic, as verified by some 
tests of convergence.
Calculations in $^{208}$Pb show that due to sizable mixing effects 
only about $50\%$ of the total isoscalar giant dipole resonance (ISGDR) 
$3\hbar\omega$ strength is located in the energy region between 20 and 25 MeV.
This is the energy region which currently is available in experiment. 
Even above 30 MeV we find about $10\%$ of the total strength.
This indicates that the current experimantal evaluations of the ISGDR centroid 
energy may significantly underestimate its value.
\end{abstract}
  
\vskip 2cm

In formal terms many interesting nuclear modes of excitation $\vert f \rangle$,
as for instance the nuclear giant resonances, 
are generated by one-body operators of the type 
${\hat f}=\sum_{\alpha \beta} f_{\alpha \beta} a^{\dag}_{\alpha} a_{\beta}$
such that 
\begin{equation}
\vert f \rangle = {\hat f} \vert 0 \rangle= 
\sum_n \langle n \vert {\hat f} \vert 0 \rangle \vert n \rangle,
\end{equation} 
The symbol $\vert 0 \rangle$ represents the ground state and $\vert n \rangle$ the
spectrum of eigenstates in a corresponding subspace.    
For such excitation modes certain global aspects of the strength function
\begin{equation}
S_{f}(E) = \sum_n S_{f}(n) \delta (E-E_n),
\end{equation}
where 
\begin{equation}
S_{f}(n) =
\vert\langle n \vert{\hat f}\vert 0 \rangle\vert^2,
\end{equation}
can thus be described in the subspace of one-particle -- one-hole (1p1h)
$(\vert 1 \rangle = a^{\dag}_p a_h \vert 0 \rangle)$
states generated by the nuclear mean field.  
In the 1p1h subspace we thus have 
$\vert n \rangle = \sum_1 c^n_1 \vert 1 \rangle$.
In general, however, such states no longer remain the exact eigenstates when
more complex configurations of the npnh-type are taken into account.
The 1p1h components of a much larger number of new eigenstates 
$\vert n \rangle$ are then spread over many more corresponding new eigenvalues.
This is a mechanism of fragmentation. 
The nuclear interaction is predominantly two-body in nature 
and thus directly couples the 1p1h states to the 2p2h ones only. Therefore, 
in practical terms it is enough [1] to diagonalize the nuclear 
Hamiltonian 
\begin{equation}
\hat H=\sum_i\epsilon_i a_i^{\dag} a_i+{1\over 4}\sum_{ij,kl}v_{ij,kl}
a_i^{\dag} a_j^{\dag} a_la_k,
\label{eq:H}
\end{equation}
in the combined space of 1p1h and 2p2h states.
In this equation the first term denotes the mean field which in the present 
work is taken as a local Woods-Saxon potential including the
Coulomb interaction. The second term is the  
residual interaction with antisymmetrized matrix elements $v_{ij,kl}$ 
and in the following discussion is represented by the density-dependent
zero-range interaction of Ref. [2].

The Hamiltonian matrix then reveals the following structure:

\begin{center}
\epsfig{file=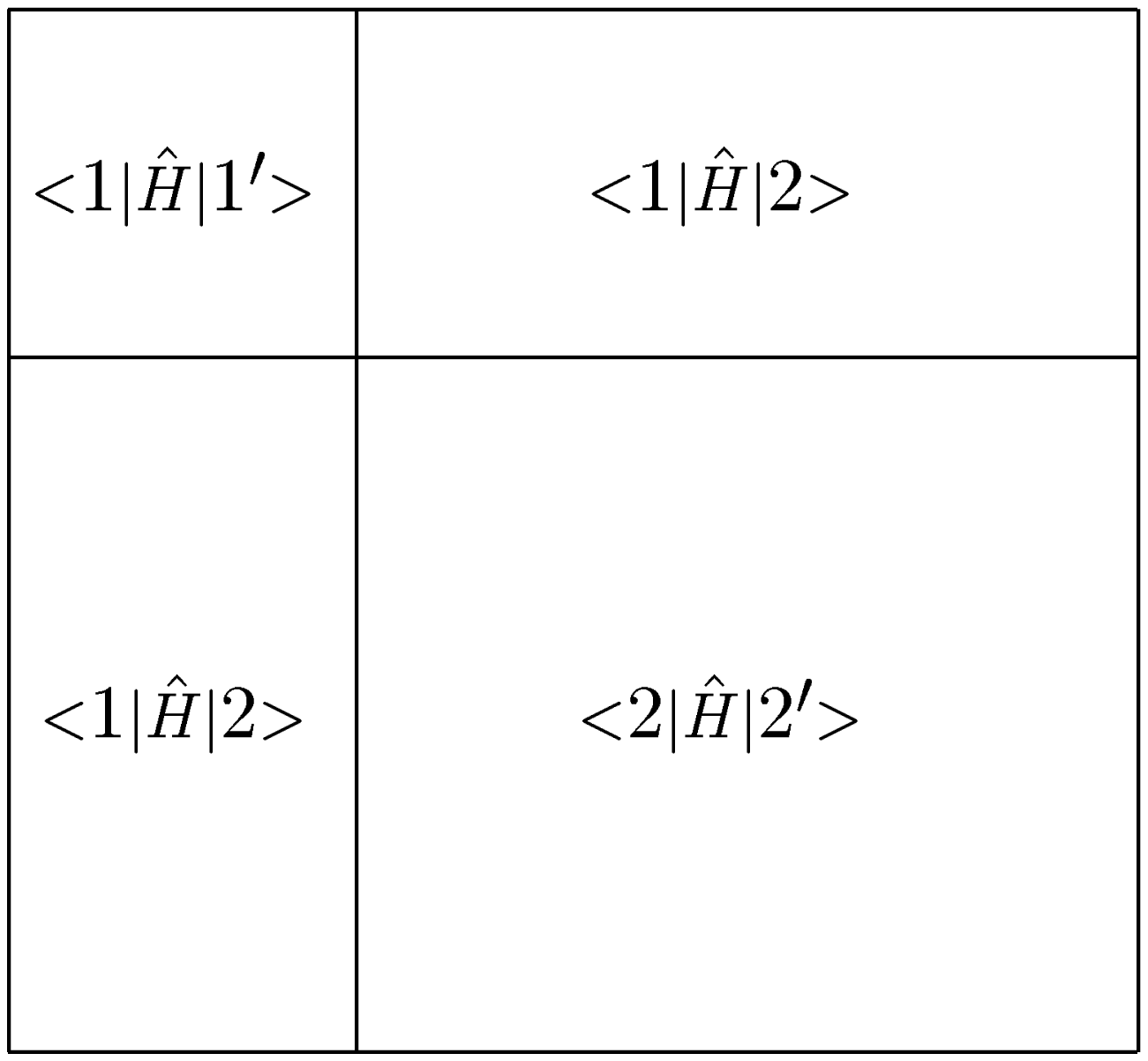,width=5cm}
\end{center}
Even after such a truncation the dimension of the above matrix 
is usually still much too large to be numerically diagonalized. 
This in particular holds true for the isoscalar giant dipole 
resonance (ISGDR) [3]. This $3\hbar\omega$ excitation is typically 
located in the energy region above 20 MeV and the number of relevant
2p2h states is of the order of $10^6$. An interesting conclusion can
however be drawn from Fig.~1 which shows the distribution of the 
coupling matrix elements $\langle 1 \vert {\hat H} \vert 2 \rangle$
between the 1p1h and 2p2h states in the $J^{\pi}=1^-$ sector of the 
$^{208}$Pb nucleus.
These states are here generated by the six mean field
shells (three above and three below the Fermi surface). As it is clearly
seen from Fig.~1 there is only a very small fraction of the coupling matrix
elements which significantly differ from zero. This seems to offer an extra
criterion for selecting the most important 2p2h states.         
Thus, by setting a finite positive threshold value $H_{th}$ one can select the
2p2h states such that 
$\vert \langle 1 \vert {\hat H} \vert 2 \rangle \vert \ge H_{th}$   
for those 1p1h states which carry the strength.
Interestingly, an explicit numerical verification shows that even a relatively
severe selection according to this prescription on average preserves 
a form of the distribution of 2p2h states. What in this connection is 
particularly important is that, as an example in Fig.~2 illustrates,
even certain high energy 2p2h states survive selection 
and thus a possibility to move the strength to higher energies is
retained. 

We now evaluate the ISGDR strength distribution using a prescription 
as described above. ISGDR is one of the most interesting nuclear excitation
modes. This partly originates from the fact that its centroid energy can 
directly be related to the nuclear compression modulus [4].  
The corresponding one-body isoscalar dipole operator reads:
\begin{equation}
f=r^3 Y_1 - \eta r Y_1,
\label{eq:ob}
\end{equation}
where $\eta = 5 \langle r^2 \rangle /3$.
The second term in this equation removes the spurious center of mass motion
component from the operator $r^3 Y_1$ [5].
The resulting $3\hbar\omega$ strength distribution in $^{208}$Pb 
on the 1p1h level is shown in Fig.~3. 
Almost all this strength is located between 20 and 25 MeV. 
This about corresponds to the energy region where the isoscalar dipole strength
can be identified in the present day experiments on $^{208}$Pb [6].  
The picture changes however significantly when mixing due to the coupling
to 2p2h states is allowed. This is illustrated in Fig.~4  which on 
the three successive panels indicates a degree of fragmentation for 
$H_{th}=0.4$, 0.3 and 0.2 MeV (from bottom to top). The number of the
corresponding 2p2h states included equals 349, 1125 and 4374, respectively.
Consistently with our previous investigations [7] a specific form
of the resulting strength distribution strongly depends on many factors and thus also on 
$H_{th}$.
However, more global characteristics, like a percentage of the total strength
in certain sufficiently large energy windows is much more stable as can be
concluded from Table 1 which lists such quantities for energies above 25 and
30 MeV, respectively. 

A reasonable convergence of those results, together
with a realistic input of the present model, provides
quite a convincing indication that one may expect about $50\%$ of the total
$J^{\pi}=1^-$ isoscalar $3\hbar\omega$ strength in the higher energy region,
above 25 MeV, i.e., in the region which is dominated by many other multipoles
and thus this portion of the strength escapes an experimental
detection. Even above 30 MeV one finds almost $10\%$ of the total
strength. The present calculations thus suggest that a recent empirical
estimation [6] of the nuclear incompressibility $(K_A=126 \pm 6$ MeV)
for $^{208}$Pb may appear much to low.
\vskip 0.5cm


This work was supported in part by Polish KBN Grant No. 2 P03B 140 10
and by the German-Polish scientific exchange program.

\vskip 2cm

{\small {\bf References:}

\begin{enumerate} \itemsep 1mm

\item[{[1]}] S.~Dro\.zd\.z, S.~Nishizaki, J.~Speth and J.~Wambach,
Phys. Rep. {\bf 197}, 1(1990)
\item[{[2]}] B.~Schwesinger and J.~Wambach, Nucl. Phys. {\bf A426},
253(1984)
\item[{[3]}] M.N.~Harakeh, Phys. Lett. {\bf 90B}, 13(1980);\\
M.N. Harakeh and A.E.L.~Dieperink, Phys. Rev. {\bf C23}, 2329(1981);\\
H.P.~Morsch {\it et al.}, Phys. Rev. {\bf C28}, 1947(1983)
\item[{[4]}] S.~Stringari, Phys. Lett. {\bf 108B}, 232(1982);\\
R.~de~Haro, S.~Krewald and J.~Speth, Phys. Rev. {\bf C26}, 1649(1982)
\item[{[5]}] N.~van~Giai and H.~Sagawa, Nucl. Phys. {\bf A371}, 1(1981)
\item[{[6]}] B.F.~Davis {\it et al.}, Phys. Rev. Lett. {\bf 79}, 609(1997) 
\item[{[7]}] S.~Dro\.zd\.z, S.~Nishizaki and J.~Wambach, 
Phys. Rev. Lett. {\bf 72}, 2839(1994);\\
A.~G\'orski and S.~Dro\.zd\.z, Acta Phys. Polonica {\bf B28},1111(1997);\\ 
S.~Dro\.zd\.z, S.~Nishizaki, J.~Speth and M.~W\'ojcik, Phys. Rev. {\bf E57},
4016(1998)

\end{enumerate}}      

\newpage
\begin{center}
{\bf FIGURE CAPTIONS}
\end{center}
{\bf Fig.~1.} Distribution of the coupling matrix elements between the 1p1h
and 2p2h states for $J^{\pi}=1^-$ sector in $^{208}$Pb.\\
{\bf Fig.~2.} Energy distribution of all the 2p2h states generated 
by the six major mean field shells (a). (b) corresponds to those 2p2h states
(2) which fulfil the condition 
$\vert \langle 1 \vert {\hat H} \vert 2 \rangle \vert \ge 0.3$ MeV.
The symbol N denotes the number of states in the bin of energy equal to
0.2 MeV.\\  
{\bf Fig.~3.} Isoscalar $3\hbar\omega$ dipole strength distribution in $^{208}$Pb
calculated in the subspace of 1p1h states.\\
{\bf Fig.~4.} Isoscalar $3\hbar\omega$ dipole strength distribution in  $^{208}$Pb 
calculated in the space of 1p1h and 2p2h states, for three different
values of $H_{th}$.\\

\begin{center}
{\bf TABLE CAPTION}
\end{center}

{\bf Table.~1.} Percentage of the total isoscalar $3\hbar\omega$ dipole strength in $^{208}$Pb
calculated in the space of 1p1h and 1p1h states in the energy region above 25 MeV and above 30 MeV,
respectively, for the three different values of $H_{th}$.
The numbers in parenthesis list the corresponding numbers of 2p2h states.\\

\end{document}